\documentclass[11pt,preprint]{aastex}

\usepackage{graphicx}

\slugcomment{}

\shorttitle{Stars, gas and dust in $z\sim 2.8$ obscured quasars}
\shortauthors{Lacy et al.}

\begin{document}


\title{The stellar, molecular gas and dust content of the host 
galaxies of two $z\sim 2.8$ dust obscured quasars.}


\author{
M.\ Lacy \altaffilmark{1}
A.O.\ Petric\altaffilmark{2},
A. Mart\'\i nez-Sansigre\altaffilmark{3,4},
S.E.\ Ridgway\altaffilmark{5},
A.\ Sajina\altaffilmark{6},  
T.\ Urrutia\altaffilmark{7}, 
D.\ Farrah\altaffilmark{8}}
\altaffiltext{1}{North American ALMA Science Center, National Radio Astronomy Observatory, 520, Edgemont Road, 
Charlottesville, VA 22903, USA}
\altaffiltext{2}{California Institute of Technology, Pasadena,CA91125, USA}
\altaffiltext{3}{Institute of Cosmology and Gravitation, University of Portsmouth, Dennis
Sciama Building, Burnaby Road, Portsmouth, PO1 3FX, UK}
\altaffiltext{4}{SEP{\i t net}, South East Physics Network, UK}
\altaffiltext{5}{NOAO, Colina El Pino s/n, Casilla 603, La Serena, Chile}
\altaffiltext{6}{Tuffs University, Medford, MA, 02155, USA}
\altaffiltext{7}{Spitzer Science Center, Caltech, Mail Code 220-6,Pasadena, CA 91125, USA}
\altaffiltext{8}{Department of Physics and Astronomy, University of Sussex, Brighton, BN1 9RH, UK}

\begin{abstract}

We present optical through radio observations of the host galaxies of 
two dust obscured, luminous quasars selected in the mid-infrared, at $z=2.62$
and $z=2.99$, including a search for CO emission.
Our limits on the CO luminosities are consistent with these
objects having masses of molecular gas $\stackrel{<}{_{\sim}} 
10^{10}M_{\odot}$, several times less than those of 
luminous submillimeter-detected galaxies (SMGs) at comparable redshifts. 
Their near-infrared spectral energy distributions (SEDs), however, imply
that these galaxies have high stellar masses ($\sim 10^{11-12}M_{\odot}$)
The relatively small reservoirs of molecular gas and low dust masses
are consistent with them being relatively mature systems at high-$z$.

\end{abstract}



\section{Introduction}

The hosts of high redshift quasars and 
radio galaxies provide an important insight into
the formation of the most massive galaxies at high redshifts. 
In particular, they can be used to study the origin of the black hole 
mass -- bulge mass/velocity dispersion 
relation (Magorrian et al.\ 1998, Gebhardt et al.\ 2000; Ferrarese 
et al.\ 2000)
and its possible explanation in terms of AGN feedback on star formation in the
host galaxy (Silk \& Rees 1998; King 2003). 
Dust obscured quasars play a key role in such models, representing the 
intermediate
stage after the black hole has begun accreting gas from its surroundings, but 
before AGN driven winds have cleared gas and dust from the
line of sight to the quasar (Fabian 1999; Li et al.\ 2006), in 
a manner similar to that proposed by Sanders et al.\ (1988).

In order to test these models, it is particularly important to build
up samples of dust-obscured quasars, especially at high redshifts, 
where an increased amount of quasar
obscuration is thought to be due to dust in the
host galaxy (e.g.\ Treister et al.\ 2009), 
rather than by a nuclear torus as in the orientation-based
unified scheme for local Seyfert and radio galaxies/quasars (Antonucci 1993).
Over the past few years, we have been constructing a sample of mid-infrared
selected, dust obscured quasars, building on the work of Lacy et al.\ (2007a). 
This sample includes $\sim 30$ objects with redshifts past the 
peak in number density of the normal quasar population at $z\sim 2$
(Richards et al.\ 2006 ).

Obscured quasars also have an useful practical advantage over
normal quasars, namely that the obscuration of the quasar nucleus by dust
allows the host galaxy of the quasar to be characterized much more
easily. Only in these obscured quasars is it
possible to obtain accurate estimates of host galaxy stellar 
luminosities and galaxy morphologies (Zakamska et al.\ 2006; 
Lacy et al.\ 2007b).

Despite their importance, obscured quasars remain a 
relatively poorly studied population. Large numbers have been 
found in the Sloan Digital Sky Survey (Reyes et al.\ 2008), but
only through mid-infrared selection has it proved
possible to find more than a handful of these objects at $z>0.8$, where
the [O{\sc iii}] emission line is redshifted out of the rest-frame optical
band (Mart\'\i nez-Sansigre et a.\ 2005; Lacy et al.\ 2007a; 
Polleta et al.\ 2006; 2009). 
Existing observations of high-$z$ dust obscured quasars
do tend to support a scenario in which a significant fraction of the dusty 
quasar population is reddened by dust in the host galaxy, rather
than a nuclear torus (Mart\'\i nez-Sansigre et al.\ 2006; Lacy et al.\ 
2007b), but key unknowns in the observational studies of these objects
remain. These include the masses of the
molecular gas reservoirs in these objects, and the stellar luminosities
of the hosts. From work that has been performed on similar 
objects, including $z\sim 2$ Ultraluminous Infrared Galaxies (ULIRGs) 
selected at 24$\mu$m with {\em Spitzer}
(Yan et al.\ 2010), many of which are AGN-dominated,
and on a few other type-2 quasars at $z>2$ (Mart\'\i nez-Sansigre et al.\
2009; Polletta et al. 2011), 
it seems that molecular gas is present in such objects, but
how the molecular gas content relates to other 
host galaxy properties is unclear.

We were thus motivated to study the CO emission from luminous
obscured quasars, since it traces the molecular gas content of the galaxy, and
hence the reservoir of molecular gas available for star formation.
In particular, the low-$J$ transitions provide the 
best tracer for the total gas mass (Papadopoulos \& Ivison 2002).
The low critical density of these transitions allows them to be
seen against the Cosmic Microwave Background 
in gas with density as low as $\sim 10^2{\rm cm^{-3}}$. 
To do this, we use the 
new capabilities of the Extended Very Large Array (EVLA). The VLA has 
a large collecting area, but has hitherto been restricted through not having
receivers in the 30-40GHz range, or a correlator capable of correlating
a wide bandwidth. The EVLA, with Ka-band receivers and the wide-band 
WIDAR correlator provides the opportunity to study these objects in the
lowest CO (1-0) transition.
When fully commissoned, the WIDAR correlator will be able to correlate an
8GHz bandwidth. It will thus be capable of measuring redshifts for galaxies
whose distances are relatively poorly constrained (for example, galaxies
with only photometric redshifts), and for galaxies whose emission features
in the optical/near-infrared are obscured by dust.
For this initial investigation, 
we used the restricted correlator bandwidth available through the Open 
Shared Risk Observation (OSRO) program and objects with well-constrained
redshifts with relatively narrow ($<5000$kms$^{-1}$) emission lines. 
The combination of these data with newly released archival  
data from the {\em Herschel}
telescope allows us to study the gas and dust content of these objects for the
first time.
We assume a cosmology with $\Omega_{M}=0.3, \Omega_{\Lambda}=0.7$ and
$H_0=70 {\rm kms^{-1}Mpc^{-1}}$.

\section{Sample selection}

Our objects were selected in the mid-infrared using the 
selection technique of Lacy et al.\ (2004), which  
finds candidate AGN and 
quasars, both obscured and unobscured, by virtue of their red mid-infrared 
continua. Our obscured quasars comprise both dust-reddened type-1 quasars
(objects with red continua, but broad permitted lines),
and type-2 quasars, selected from the {\em Spitzer} SWIRE 
survey (Lonsdale et al.\ 2003) and the 
Extragalactic First Look Survey (XFLS) 
(Lacy et al.\ 2005; Fadda et al.\ 2006). 

Optical spectra of the first 77 objects in this mid-infrared selected 
sample were published in Lacy et al.\ (2007a).
A further $\sim 700$ 
spectra have now been obtained (Lacy et al.\ 2012, in preparation). 
From this list we picked out a pilot sample of 
two obscured quasars  for follow-up 
with the EVLA in OSRO mode, according to the 
following criteria: $z>2.5$, $16<{\rm RA}<20$ (to ensure good RAs for 
night time summer observing) and 
optical spectroscopic redshifts from multiple emission lines
(Figure 1). Unlike most other samples of quasars selected for CO follow-up, 
our objects were not pre-selected for bright far-infrared emission.

The high redshifts of our sources mean that the available optical spectra
sample the rest-frame UV. Our working definition of a type-2
quasar as an object which lacks broad permitted 
lines in the rest-frame optical. As only 
XFLS171419.97+602724.8 has a rest-frame optical spectrum 
(Figure 2), we are only sure of this one as being a
type-2 quasar by our definition. However, the spectral energy distribution 
(SED) of 
SW161117.30+541759.2 suggests it is dominated by starlight in the 
rest-frame optical. The spectra 
are discussed in more detail in Section 3.

\section{Optical and infrared observations}

XFLS171419.97+602724.8 has had its optical spectrum 
previously published in Lacy et al.\ (2007a)
(but reproduced here for ease of comparison with the near-infared 
spectrum presented here). 
The spectrum of SW161117.30+541759.2  was taken using Hectospec on the MMT
during June 2008, we used the 270 groove per mm grating, resulting 
in a spectral resolution
of $\approx 0.5$nm. Approximately 130 fibers were placed on candidate 
AGN in each setting. 
The field was observed for for $3\times 1800$s. 

\begin{figure}
\plotone{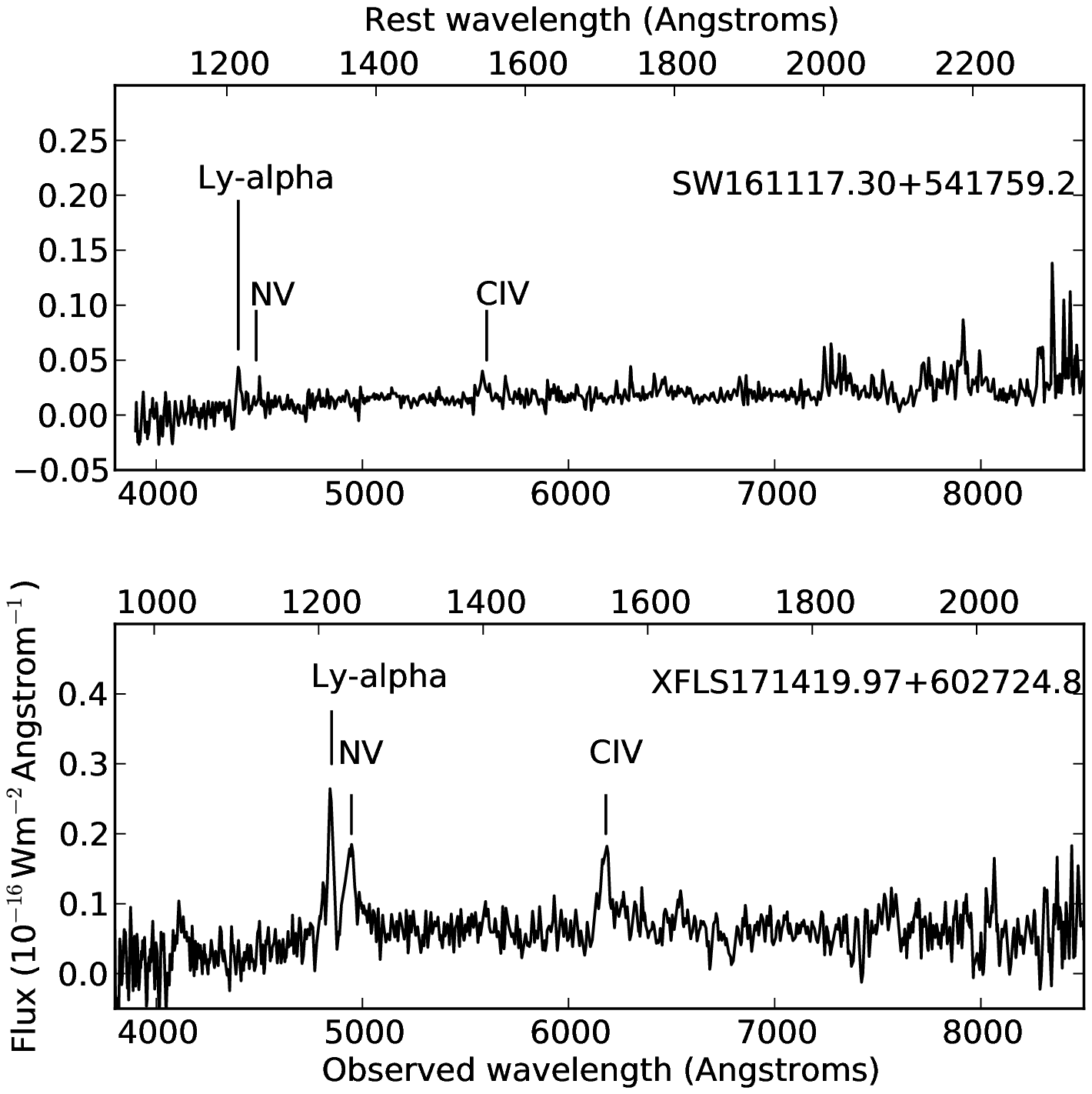}
\figcaption{Optical (rest-frame UV) spectra of our two high redshift obscured
quasars.}
\end{figure}

\begin{figure}
\plotone{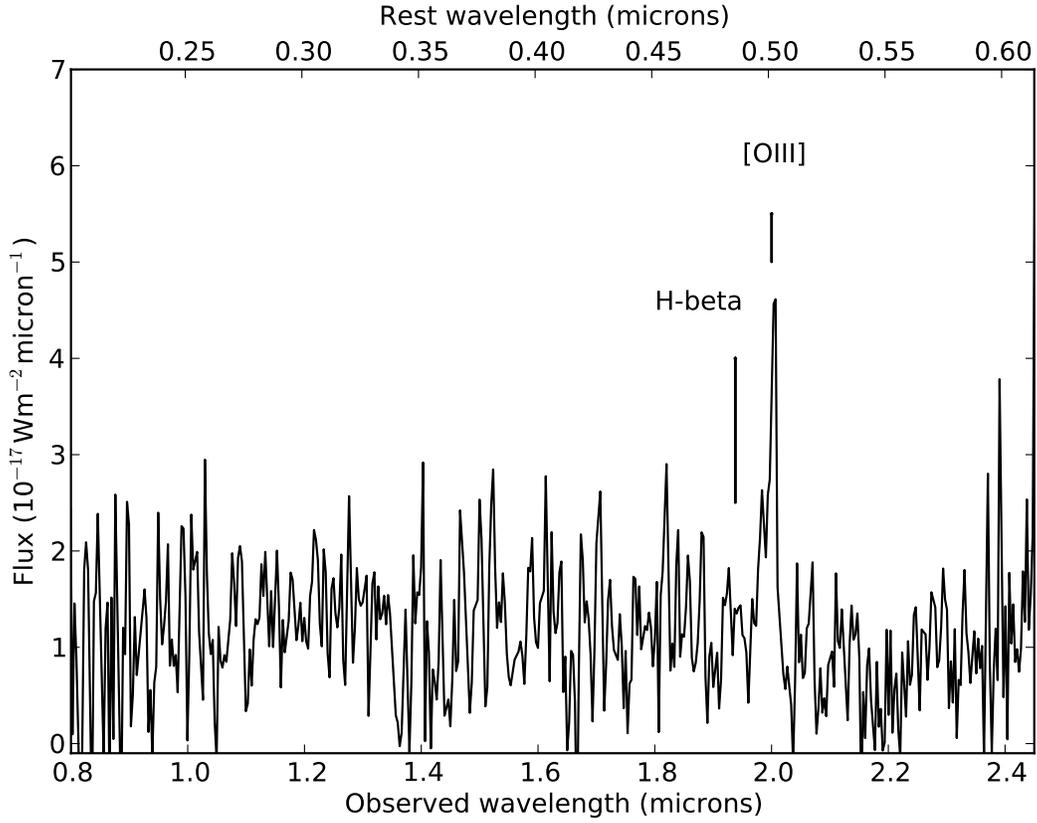}
\figcaption{Near-infrared (rest-frame optical) spectrum of XFLSJ171419.97+602724.8.}
\end{figure}

\begin{table}
\caption{Redshifts and photometry}
{\scriptsize

\begin{tabular}{lcc}
Object &SW161117.30+541759.2 &XFLS171419.97+602724.8\\\hline
Redshift  &   2.618  & 2.986\\
$u$ (AB)  &   $>$24.2&$>$22.0\\
$g$ (AB)  &24.1&     21.9\\
$r$ (AB)  &$>$23.7&21.5\\
$i$ (AB)  &23.1&21.0\\
$z$ (AB)  &$>$22.2&$>$20.5\\
$J$ (Vega)&21.5 & - \\
$H$ (Vega)& -   & 18.3\\
$K$ (Vega)& 19.4& - \\
3.6$\mu$m ($\mu$Jy)&43.4&178\\
 4.5$\mu$m ($\mu $Jy)&86.3&248\\
 5.8$\mu$m ($\mu $Jy)&204&495\\
 8.0$\mu$m ($\mu $Jy)&279&1080\\
 24$\mu$m ($\mu$Jy)&1155&5600\\
 70$\mu$m ($\mu $Jy)&$<$15000&26000\\
 160$\mu$m ($\mu $Jy)&$<$60000&59000\\
 250$\mu$m ($\mu $Jy)&$<$9000&$<$18000\\
 350$\mu$m ($\mu $Jy)&$<$9000&$<$15000\\
1.4GHz ($\mu $Jy)&$<$135& 1230\\
610MHz ($\mu $Jy)&$<$200&2160\\
\end{tabular}
}
\end{table}

The data were analysed using 
software based on the {\sc hsred} package of R.\ Cool,
with a small modification to improve the sky subtraction.  
Details are given in Lacy et al.\ (2012).
The resulting spectrum is shown in Figure 1 (upper panel).
In most cases our sky subtraction technique works well, particularly for the 
discrete sky lines. However, fiber spectra of faint objects 
are more vulnerable to 
artifacts due to poor sky subtraction than longslit spectra. We therefore
also checked that the redshift was consistent with the continuum SED, 
especially in the area around the Balmer break in the
observed near-infrared (Section 5.2).

A near infrared spectrum of 
XFLS171419.97+602724.8 was obtained using SpeX (Rayner et al.\ 2003) 
at the Infrared Telescope Facility (IRTF)
on 2007-06-18 (Figure 2). SpeX was used in low dispersion (prism) mode with a 0.5 arcsec slit, 
resulting in a resolving power $\approx 150$. The object was nodded between 
two positions along the slit in an ABBABBA... pattern, with 2 min integration times at 
each nod position, for a total of 1hr of 
integration time on the object. The airmass was 1.6, and the slit aligned along the parallactic angle. 
The atmospheric transmission and flux was calibrated on the nearby A0 star HD172728. An $H$-band image was obtained
using SpeX in imaging mode on 2007-06-19 using $7\times 2$min exposures at an airmass of 1.7, 
calibrated on the standard P138-C (Persson et al.\ 1992). 
XFLS171419.97+602724.8 also has a spectrum in the {\em Spitzer} archive, 
Astronomical Observation 
Request number 24191744 from Program 40539 (Wu et al.\ 2010), taken with the
Infrared Spectrograph (IRS). The spectrum,
which runs from 5-38$\mu$m,  
was extracted using the {\em Spice} tool from the {\em Spitzer} Science 
Center\footnote{http://ssc.spitzer.caltech.edu/dataanalysistools/tools/spice}. The spectrum itself is featureless, 
(in particular, no Polycyclic Aromatic Hydrocarbon [PAH] 
features are visible), but 
the overall shape constrains the mid-infrared SED very well (see section 5.2).

Optical magnitudes for these objects were obtained from the Sloan Digital
Sky Survey (Adelman-McCarthy et al.\ 2008), from 
Gonz\'alez-Solares et al.\ (2011) and from 
Fadda et al.\ (2004). Near infrared magnitudes and images in $J$- and 
$K$-band for SW161117.30+541759.2 were obtained from the 
UKIRT Infrared Deep Survey (UKIDSS) Deep Extragalactic Survey. 
Infrared fluxes from {\em Spitzer} at 3.6-160$\mu$m
were obtained from SWIRE (data release 2) and the XFLS (Lacy et al.\
2005; Fadda et al.\ 2006; Frayer et al.\ 2006). 

{\em Herschel} SPIRE limits at 250 and 350$\mu$m were obtained from the 
Level 2 {\em Herschel} archival products originally taken 
as part of the Hermes project (Oliver et al.\ 2010), 
using aperture photometry, assuming beam areas from the SPIRE beam release 
note version 1 (Sibthorpe et al.\ 2010), and using aperture corrections 
derived based on those beams. Radio 
fluxes/limits at 1.4GHz are from Condon et al.\ (2002) and Ciliegi et al.\ (1999) and at 610MHz from Garn et al.\ (2007, 2008). The available photometry for
our objects is listed in Table 1.

\section{EVLA Observations}

Our quasars were observed with the EVLA in 
OSRO mode as program AL744. Five objects in total were observed, however,
three were later established to be highly-extincted starbursts or quasars
at lower redshift and
have been removed from the analysis in this paper.
Our two remaining objects were
observed in Ka-band. 
Only a single intermediate frequency could be tuned low enough
to reach the redshifted CO(1-0) line in these objects, and 
hence the available bandwidth
was only 128MHz (see Table 2). We refer to the IF that was not tuned to the 
CO(1-0) line frequency as the ``off-tuned'' IF. 
All observations used a 2MHz channel width ($\approx 20$kms$^{-1}$). 

The available bandwidth of 
128MHz in Ka-band (corresponding to $\Delta z=0.016$ at $z=2.8$, or
a velocity range of 1270kms$^{-1}$) 
required us to use redshifts
as close as possible to the systemic velocity of the galaxies. As our optical
spectra are typically of modest signal-to-noise, with the brightest
lines being high ionization species which can often be in outflows, we 
chose to estimate the systemic redshift from the average velocity of the 
lowest ionization, highest signal-to-noise lines.

Both objects were observed in 3hr scheduling blocks, with total times on 
source $\approx$2.1hr (exact observation times and RMS noises obtained 
are given in Table 2). 
Phase calibration was performed on a nearby calibrator, 
(J1620+4901 for SW161117.30+541759.2, J1746+6226 for
XFLS171419.97+602724.8),
using on-source scans of six minutes between each observation of the 
calibrator. Flux calibration was performed on 3C48. The initial flux
calibrations failed due to an error in the execution of the scheduling block which
led to the last scan of the block being dropped. They were thus repeated on 2010-09-13, 
however, only a
K-band flux calibration on J1746+6226
proved successful, for unknown reasons. 
The fluxes of J1746+6226
in Ka band were extrapolated from that in K-band using a spectral index
between 8.4GHz and 30GHz obtained from previous observations 
(Patnaik et al.\ 1992; Healey et al.\ 2007; Lowe et al.\ 2007) 
listed in the 
NASA Extragalactic 
Database (NED). For J1620+4901 we made use of our observations in the off-tuned
IF (selected to be centered at 37GHz for all the Ka-band observations)
to obtain the ratio of the fluxes of J1620+4901 and
J1746+6226 at 37GHz, then extrapolated the flux of J1620+4901 to the redshifted
CO line frequency at 31.9GHz using
its spectral index from NED. Consequently, we estimate that there is an unusually high
uncertainty in the flux calibration of the Ka-band observations of $\approx 20$\%, based on the uncertainty in the spectral index, which may have varied
between the epoch of the archival observations in NED and the present.

The data were analysed in CASA using standard procedures\footnote{see
http://casaguides.nrao.edu/index.php?title=EVLA\_Tutorials}. The spectra were
then extracted using a beam-weighted extraction centered on the mid-infared
position of the sources. The noise per channel was estimated by measuring the
RMS on each image plane of the cube.

\begin{figure}

\centering
\includegraphics[width=3.0in]{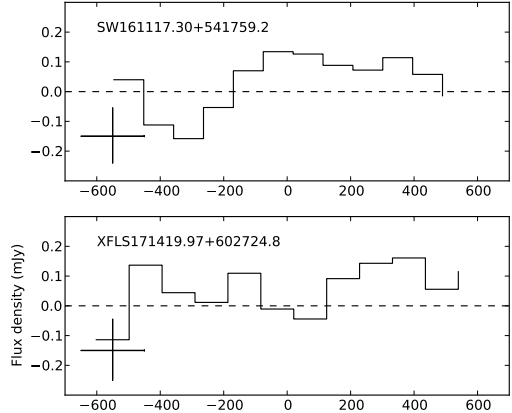}

\caption{CO (1-0) spectra of our two objects, smoothed by ten 
channels ($\approx 100 {\rm kms^{-1}}$) and sampled every five. In 
the case where we have detected a line at $>3\sigma$ (XFLS171216.48+595711.0)
the horizontal solid line shows the velocity range integrated over to obtain the line flux. 
The crosses on the lower left of each plot indicate the flux density error on each point,
and the width of the boxcar smoothing used.} 

\end{figure}

\begin{table*}
\caption{EVLA observations}
{\scriptsize
\begin{tabular}{lcc}
Object          & SW161117.30+541759.2 &XFLS171419.97+602724.8$^{*}$\\\hline
Frequency (GHz) &31.85977 &28.91900\\
Band            & Ka      & Ka     \\
Bandwidth$^{\dag}$ (MHz) & 128     & 128  \\
Date Observed   & 2010-06-28&2010-07-13,16\\
N$_{\rm ants}^{\ddag}$ & 17 & 20\\
Integration time (mins)& 123 & 2$\times$126\\
RMS noise (mJy) & 0.33 & 0.36\\
$I_{\rm 1-0}$ (mJykms$^{-1}$)& $<$110 & $<$90\\
$L^{'}_{CO}$ $(10^{10} {\rm Kkms^{-1}pc^{-2}}$ & $<$3.5 & $<$3.5\\\hline
\end{tabular}

\noindent
$^{\dag}$Although the OSRO bandwidth is 256MHz, only a single intermediate frequency 
of 128MHz bandwidth could be tuned below 33GHz.
$^{\ddag}$Number of antennas used in mapping. (Some antennas were yet to be fitted with Ka-band receivers.)
$^{*}$ The scheduling block for XFLS171419.97+602724.8 was observed twice, on 2010-07-13 and 2010-07-16, however, the second set of data was significantly noisier. The cubes from each observation were combined with relative weights of 0.73:0.27 for the 2010-07-13 and 2010-07-16 observations, respectively.
}
\end{table*}

\section{Results}

\subsection{CO luminosities and masses}

Our $3\sigma$ limits on the CO(1-0) luminosities are shown in Table 2. 
SW161117.30+541759.2 is formally
detected at the $\sim 2\sigma$ level, but further observations will be
needed to confirm this. There is the 
possiblity of a small amount of continuum contamination in 
XFLS171419.97+602724.8 - if the spectral index between 610 and 1400MHz is
extrapolated to 30GHz, the object would have a flux density of 156$\mu$Jy,
compared to a $2\sigma$ continuum limit of $\approx 100\mu$Jy. However, the 
spectral index most likely steepens at high frequencies, so the true continuum
flux is likely to be $<<100\mu$Jy.

There is the possibility
that a significant amount of line flux is missing due to the 
restricted bandwidth used, only repeat observations with the full bandwidth 
of the
WIDAR correlator will ensure that this is not the case. However, none of our  
observations have obvious wings of lines cut off at the edge of the bandwidth, 
suggesting that any missed flux is unlikely to be very large.

The value of the conversion factor, 
$\alpha_{CO}$ between the velocity integrated CO brightness temperature 
luminosity, $L^{'}$ (in units of Kkms$^{-1}$pc$^{2}$), and molecular 
gas mass in solar masses, $M(H_2)$,
has been a source of considerable controversy (see Ivison et al.\ 2011 for
a recent discussion). CO is almost 
always optically thick, and observed values of 
$X_{CO}$ range from $\sim 5$ in giant molecular clouds in the Milky Way, to $0.3-1.3$ in the
interstellar medium of Ultraluminous Infrared Galaxies. For SMGs, a value of 
$\alpha_{CO}=0.8$ is typically assumed, but is highly uncertain. We
assume a value of $\alpha_{CO}=0.8$ for ease of comparison with other work.

\subsection{Fitting the Spectral Energy Distributions}

The Spectral Energy Distributions (SEDs) of our objects were fit using the 
model
for AGN and starburst emission employed by Lacy et al.\ (2007b) to model low 
redshift type-2 quasars, and  
detailed in Sajina et al.\ (2006) and Hiner et al.\ (2009). For the AGN, we 
used
a power-law with a cutoff at the
dust sublimation temperature, assumed to be 1500K, and a long wavelength cutoff
at $\approx 20\mu$m, corresponding to a typical nuclear torus SED.
For SW161117.30+541759.2 we also added a ``very hot'' AGN 
blackbody component of 
dust as detailed in Hiner et al.. 
The warm and cold dust components were
modeled as described by Sajina et al.\ (2006), the warm component being a power-law 
component with a cutoff at long wavelengths to simulate very small grain emission, and
the cold component a modified blackbody. We tested three different models for 
the stellar populations. Single stellar population (SSP) models from Maraston (2005)
aged 100Myr, 600Myr and 1Gyr, reddened by an SMC extinction law (Pei 1992) with 
variable optical depth in $V$-band, $\tau_V$, and a dual
population model consisting of a 100Myr and 1Gyr SSP, with a variable ratio of young to 
old stellar population ($f_{\rm YSP}$), with no reddening. The models were
deliberately kept simple to reduce the number of free parameters in the fit.
We estimated a rest-frame
$H$-band flux based on the stellar component alone. This allows us to compare directly
with the results of Hainline et al.\ 
(2010) who perform similar fits to the SEDs of submillimeter selected galaxies (SMGs). 
The best fits are shown in Figure 4 and the results in Table 3.

In the far-infrared SW161117.30+541759.2 lacks detection in either 
{\em Spitzer} or SPIRE data
longward of 24$\mu$m, so we placed a limit on the far-infrared luminosity by
fitting the maximal warm plus cold dust combination, constrained such that the warm 
component fit smoothly onto the cold (Sajina et al.\ 2006). 
XFLS171419.97+602724.8 required a relatively high ``cold'' dust temperature (
$\approx 135$K) to fit
the {\em Spitzer} 70$\mu$m and 160$\mu$m detections and the SPIRE 250$\mu$m limit. The high
dust temperature, combined with the lack of PAH detection in the IRS spectrum
of XFLS171419.97+602724.8, make it unlikely that the dust is 
heated by normal star formation in this object. If the AGN is 
responsible, however, it 
requires AGN heating to dominate on scales beyond 
the usual pc-sized torus in these objects (see e.g. Section 3.5 of Mart\'\i nez-Sansigre et al.\ 2009). 

We have estimated dust masses or limits using the prescription of Hildebrand (1983), assuming a temperature
of 50K for SW161117.30+541759.2 and the fitted 
135K for XFLS171419.97+602724.8 (Table 3). Assuming a gas-to-dust ratio similar to the 
Milky Way ($\approx 120$; Stevens et al.\ 2005) or the lower value $\approx 54$ more typical
for SMGs (Kov\'{a}cs et al.\ 2006), we find that the limits on our gas masses are consistent with
either of these ratios.

\begin{figure}

\centering
\includegraphics[width=4in]{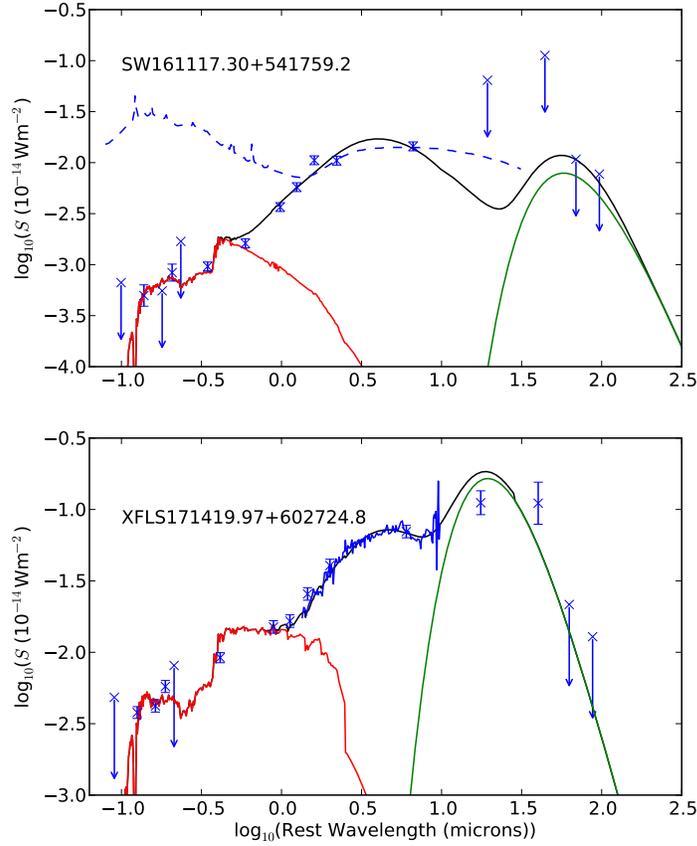}

\caption{Fits to the UV through mid-infrared spectral energy distributions
of our sources. $S = \nu S_{\nu}$ is the measured flux. Red lines 
highlight the stellar component and green lines the cold dust component.
black lines are the sum of all the model components. The blue dashed 
line on the top plot (SW161117.30+541759.2) is
a normal quasar SED matched approximately to the mid-infrared flux, and
the blue line in the plot of XFLS171419.97+602724.8 is the IRS spectrum.}
\end{figure}

\begin{figure}

\centering
\includegraphics[width=3.0in]{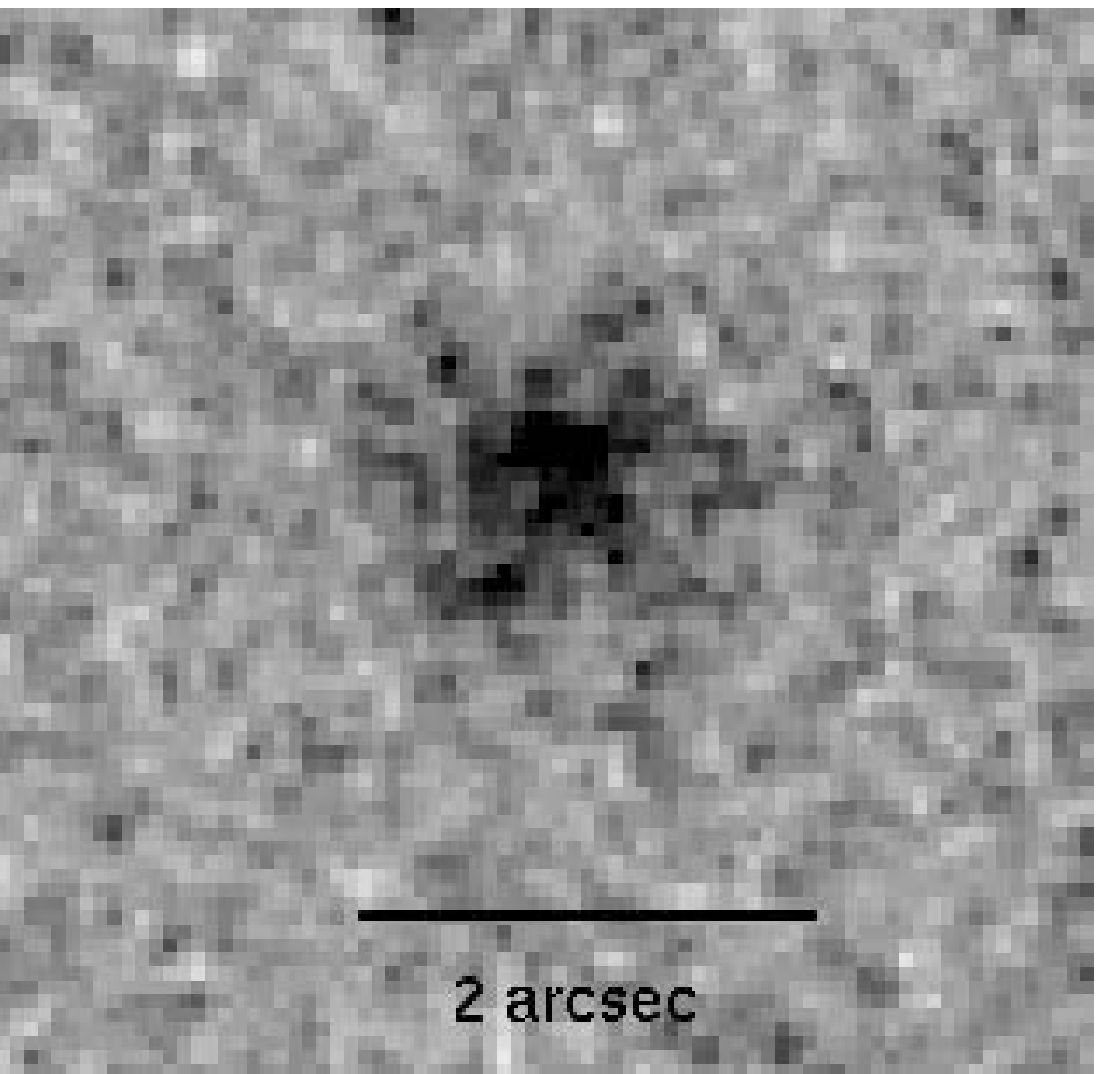}

\caption{XFLS171419.97+602724.8 in $H$-band from IRTF imaging. The object is clearly resolved in 
0.7 arcsec seeing.}

\end{figure}

\subsection{Star formation rates}

We have three ways to estimate, or constrain, the star formation rates 
(SFRs) in our 
host galaxies, though all three are vulnerable to contamination by AGN-related
emission.
The restframe UV emission, although possibly 
contaminated by scattered light of nebular continuum from the 
AGN, provides one measure, the far-infrared emission (or limits, in most cases) provides another
constraint, and the radio emission a third. 
Despite the likelihood of contamination by the AGN of one or more 
measures, we can at least investigate whether the three estimates are self-consistent. For the
star formation rate estimate based on the UV flux, we assume a 
conversion of $SFR_{UV} = L_{UV}/1.4\times 10^{21}{\rm WHz^{-1}} M_{\odot}{\rm yr^{-1}}$,
where we have evaluated $L_{UV}$ at 1530\AA. We also derive an estimate
corrected for dust extinction using the fitted $\tau_V$ and 
the SMC extinction law.
For the star formation rate based on far-infrared 
($SFR_{IR}$) emission, we have used the Kennicutt (1998) conversion:$SFR_{IR}= 4.4\times 10^{-37}
L_{FIR} ({\rm W}) M_{\odot}{\rm yr^{-1}} $. For the radio 
($SFR_R$) emission, we have used the estimate from Bell 
(2003): $SFR_R=5.5\times 10^{-22}L_{1.4GHz}M_{\odot}{\rm yr^{-1}}$,
assuming a standard radio spectral index of $\alpha_R=-0.7$, where radio 
flux density, $S_{\nu}\propto \nu^{-\alpha_R}$.

These estimates are shown in Table 3. SW161117.30+541759.2 has the 
lowest SFR, with 
no radio or far-infrared detection and only a moderate UV luminosity. 
The SFR from the 
UV and infrared/radio disagree for XFLS171419.97+602724.8, the UV value 
$\approx 130M_{\odot}{\rm yr^{-1}}$ compares with infrared and radio 
luminosities corresponding to 4000-5000$M_{\odot}{\rm yr^{-1}}$. The 
latter values seem anomalously high, this, coupled with the very high dust
temperature (135K) of the SED fit to this object, and the lack of PAH features 
in the IRS spectrum suggest that the UV estimate is closer to the 
truth, and that the FIR luminosity is heavily dominated by the AGN. 
Table 3 highlights the difficulty in estimating 
star formation rates at any wavelength in the presence of a powerful AGN.

\subsection{Stellar luminosities}

Stellar luminosity estimates are also vulnerable to contamination by AGN emission, both direct
and scattered. Direct AGN emission is typically seen as a hot near/mid-infrared component
dominated by dust heated by the AGN. Scattered light is usually blue, from
electron or dust scattering, so is typically weak in the near-infrared. 
Although radio-quiet obscured quasars have not been studied in detail to assess the
AGN light contribution to the rest-frame near-infrared emission, some
work has been performed on high redshift radio galaxies. 
Leyshon \& Eales (1998) and Leyshon, Dunlop \& Eales (1999) show that the majority 
high redshift radio galaxies have only weak
polarization in the near-infrared, indicating little scattered light at those wavelengths.
Seymour et al.\ (2008) and de Breuck et al.\ (2010) studied high redshift radio galaxies
with {\em Spitzer}, and demonstrated the practicality of separating the AGN and stellar emission
in such objects by modeling the AGN emission and fitting a simple stellar SED, in a similar
manner to the fitting done in this paper. 

One caveat that needs to be applied to our analysis is that 
our quasars were selected on the basis of a detection in all four IRAC bands in the 
XFLS and SWIRE catalogs. The XFLS catalog, in particular, has relatively bright flux 
density limits of 20,25,100 and 100 $\mu$Jy in [3.6],[4.5],[5.8] and [8.0], respectively 
(Lacy et al.\ 2005). The situation for the deeper SWIRE data is better, with 5$\sigma$
limits of  7, 10, 28, 33$\mu$Jy 
in [3.6],[4.5],[5.8] and [8.0], respectively (Lonsdale et al.\ 2003). 
Nevertheless, in both samples there may be objects missing, particularly
at high redshifts, because their
rest-frame near-infrared emission (which is typically 
dominated by starlight from the
host, rather than thermal emission from the AGN) is not bright enough to be
detected in all four IRAC bands.

SW161117.30+541759.2 is well fit by a 100Myr old
stellar population, with a modest amount of reddening. 
XFLS171419.97+602724.8 seems
to require a superposition of an old (1Gyr) and young (100Myr)  
stellar population to account for a significant Balmer break while
retaining a blue UV continuum.

\subsection{Host galaxy morphologies}

A major observational advantage to studying heavily obscured quasars is the ability to image the host free of the bright nuclear light. The host of SW161117.30+541759.2 is only marginally resolved in ground-based seeing, but 
the host of XFLS171419.97+602724.8 is clearly resolved in an $H$-band IRTF 
image in 0.7-arcsec seeing (Figure 5).

\begin{table*}
\caption{Derived quantities}
{\scriptsize
\begin{tabular}{lcc}
Object & SW161117.30+541759.2 &XFLS171419.97+602724.8\\\hline
Best model    & 100Myr SSP  & Dual 100Myr\&1Gyr\\
$f_{\rm stars}$& 0.08& 0.47\\
$\tau_V$ (SSP model)&0.34 & - \\
$f_{\rm YSP}$ (Dual model)& - &0.13\\
$M_H$ & -25.3 & -29.1\\
log$_{10}( L_{AGN} (L_{\odot}))$&12.6 & 13.4\\
log$_{10}(L_{FIR} (L_{\odot}))$&$<$12.3 & 13.3\\
$T_{\rm cold\;dust}$ (K) & [50] & 135$\pm$10\\
$\chi^2/DOF$ & 5.0 & 4.1\\
SFR (UV) (uncorrected for extinction)&9.1 & 130\\
SFR (UV) (corrected for extinction)& 31   & 130\\
SFR (FIR)& $<$350 & 4000\\
SFR (Radio)& $<$390 & 5130\\
Molecular gas mass (($10^{10}M_{\odot}$,$\alpha_{\rm CO}=0.8$) &$<$2.8&$<2.8$\\
log10($M_{\rm dust})$ &$<$7.9 & 7.0\\
log10($M_{\rm stars})$ & 10.9 & 12.4\\
log$_{10}(M_{BH}/M_{\odot})$&$>$8.1 &$>$8.9\\\hline
\end{tabular}

\noindent
Notes: $f_{\rm stars}$ is the fraction of starlight in the model in rest-frame $H$-band, $\tau_V$
is the optical depth in $V$-band and $f_{\rm YSP}$ is the fractional contribution of the 
young stellar population in the dual stellar population model. The limit
on $M_{\rm BH}$ is calculated assuming the quasar is accreting at the Eddington Rate.}

\end{table*}

\section{Discussion}

\subsection{The AGN nature of these objects}


The best spectroscopic analogues with similar redshifts as 
these objects are high redshift radio galaxies (McCarthy 1993), and type-2 
quasars detected
in X-ray surveys (e.g.\ Norman et al.\ 2002; Silverman et al.\ 2010). 
The Ly$\alpha$ lines from the radio galaxies
are, however typically much stronger relative to the UV emission lines, 
consistent with the large haloes of ionized gas around these objects being
affected by the presence of the radio jets (Reuland et al.\ 2007). 
These objects also have similar UV spectra to those of low 
redshift Seyfert-2 galaxies (e.g.\ Kriss et al.\ 1992; Kraemer et al.\ 1994),
and AGN selected through Lyman-break techniques (Hainline et al.\ 2010b). 
Thus, by analogy with known AGN both at high and low redshift we feel 
secure in assuming that the presence of high ionization, high equivalent 
width UV emission lines is a reliable tracer of AGN activity. 

\begin{figure*}

\plotone{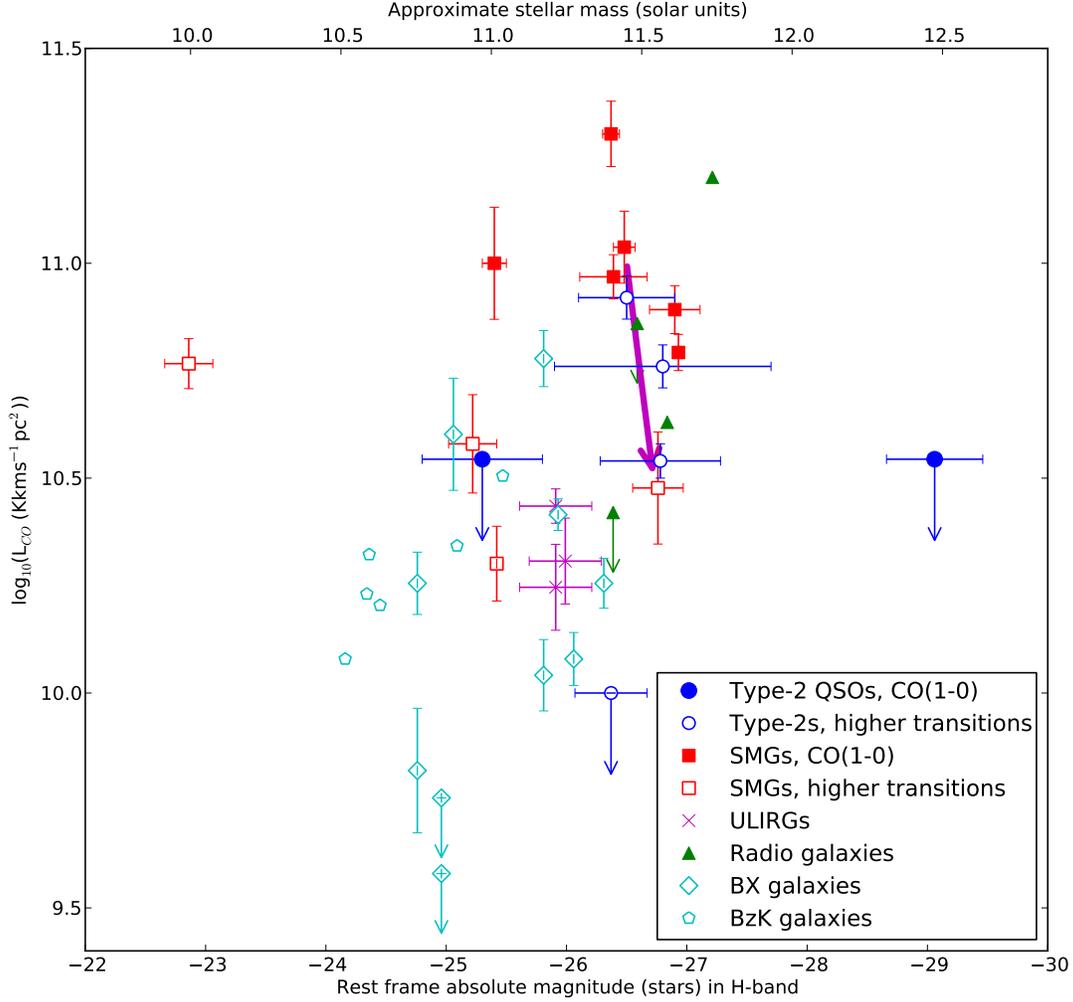}

\caption{\scriptsize CO luminosity plotted against rest-frame absolute magnitude in H-band
for the stellar component, after subtraction of the AGN model. Filled symbols
denote objects with detections or limits in CO (1-0), open symbols or
crosses those with detections in higher order CO transitions, typically
CO(3-2) 
(and therefore possibly as much as 0.3dex lower than the CO (1-0) luminosity, 
Ivison et al.\ 2011). 
The filled
blue circles are our obscured quasars with CO luminosities from CO(1-0), the
open blue circles the two $z\sim 3.5$ type-2 quasars of Polletta et al.\ 
(2011) and the two radio-intermediate type-2 quasars from 
Mart\'nez-Sansigre et al.\ (2009 and H.\ Schumacher, 
personal communication). The filled red squares are 
SMGs from Hainline et al.\
(2010a) and Ivison et al.\ (2011) with CO(1-0) measurements from either 
Hainline et al.\ (2006) or Ivison 
et al.\ (2010, 2011). The open red quasars are SMGs with measurements only in 
higher CO transitions from Greve et al.\ (2005),
Riechers et al.\ (2010) and Coppin et al.\ (2010b)
(with the $H$-magnitudes from Hainline et al.\ [2010a] or Wardlow et al.\ [2010]).
The green triangles 
are radio galaxies: 4C60.07 ($z=3.799$) with 
CO(1-0) from Greve et al.\ (2004), and TN J0924-2201 ($z=5.2$) with the
highly-luminous CO(1-0) line from Kamer et al.\ (2005), and two limits
from Emonts et al.\ (2011) (MRC0943-242 and MRC2104-242), all with $H$-band
luminosities from de Breuck et al.\ (2010). 
The magenta crosses are $z\sim 2$ ULIRGs 
from Yan et al.\ (2010) with AGN-like SEDs and for which sufficient 
photometry existed to 
estimate stellar luminosities (MIPS15494, MIPS16059, MIPS16080).
Cyan diamonds are $z\approx 2-2.5$ 
blue excess galaxies from Tacconi et al.\ (2010), and 
the cyan pentagons $z\approx 1.5$ galaxies selected according to the 
BzK technique from Daddi et al.\ (2010). Note that the Riechers et al., 
Tacconi et al.\
and Daddi et al.\ H-band absolute magnitudes have been calculated from their
estimated stellar masses given in their respective papers, assuming a 
rest-frame $H$-band mass-to-light ratio of 0.33, appropriate for a young 
stellar population (Borys et al.\ 2005), and consistent with our own fitting.
This approximate mass-to-light ratio has also 
been used to scale the top axis in stellar mass.
The large magenta arrow shows the locus of an object starting near the 
center of the SMG distribution and converting 0.5dex 
($\approx 7\times 10^{10}M_{\odot}$) of its molecular
gas to stars, assuming 100\% efficiency.}

\end{figure*}

\subsection{Comparison to CO measurements of normal quasars}

Few CO (1-0) measurements of normal quasars at similar redshifts have been published.
Riechers et al.\ (2006) measured CO(1-0) of three $z\sim 4$ quasars using the Greenbank
and Effelsberg telescopes. Their luminosities are significantly higher than those
measured for dusty quasars in this paper ($L_{CO}\approx 5\times 10^{10} {\rm Kkms^{-1}pc^2}$) , 
but the quasars are intrinsically more
luminous, and were pre-selected for having ultraluminous infrared emission, so form a
biased sample. Similarly, Coppin et al.\ (2008) detected CO emission in  
5/6 optically-luminous, submm detected quasars at $z\approx 2$, with 
luminosities of $L_{CO}\approx 1.2-8.3\times 10^{10} {\rm Kkms^{-1}pc^2}$.
At higher redshifts, Wang et al.\ (2010) find slightly lower CO luminosities,
$L_{CO}\approx 0.8-3\times 10^{10} {\rm Kkms^{-1}pc^2}$, in a sample of eight
$z\sim 6$ quasars, again pre-selected to have submm continuum detections.
At present it is therefore difficult to say whether the dusty quasars
in this paper have significantly more or less molecular gas 
than their unobscured counterparts, since ours are
unbiased towards bright far-infrared luminosities (but
see also Klamer et al.\ 2005). 
This situation is likely 
to change rapidly, however, as the EVLA and ALMA are
fully commissioned over the next 2-3 years.

\subsection{Dusty quasars and the evolutionary paradigms for massive galaxies at high redshifts}

Our understanding of galaxy evolution would be greatly enhanced if we
could compare the gas content, stellar masses, 
star formation
rates and space densities in unbiassed samples of different types 
of massive, high 
redshift galaxies, in order to constrain evolutionary scenarios. 
This would include SMGs, warm ULIRGs and obscured quasars selected
in the mid-infrared or hard X-ray, radio galaxies, normal quasars and, if 
they exist, the quiescent equivalents of each class. 
Selection effects on current samples of 
these populations are strong, however, making it difficult
to perform a definitive comparison. In our quasar hosts, 
there is a selection against
objects with low stellar mass due to the requirement for detection in 
all IRAC bands, but no pre-selection for high dust masses and star formation 
rates based on mm/submm continuum detections. 
In the samples of {\em Spitzer} selected ULIRGs
of Yan et al.\ (2010) and type-2 quasars of Mart\'\i nez-Sansigre et al.\
(2009), a continuum detection at 1.2mm
(i.e. a high star formation rate and dust mass) was a requirement for 
follow-up observations for CO. The SMGs with CO (1-0) measurements
are also restricted to a fairly small sample at present, though this 
situation is likely to change in the next few months as EVLA commissioning 
progresses (e.g.\ Ivison et al.\ 2011). 
Nevertheless, we can attempt to piece together a consistent picture with 
what is currently available.

Figure 6 shows our quasar host galaxies compared to other high redshift
objects (including other type-2 quasars) 
in the CO-luminosity -- stellar luminosity (rest-frame $H$-band
absolute magnitude) plane. Assuming similar
conversions between CO luminosity and CO mass, and stellar luminosity and
stellar mass for the different populations shown here, we can begin
to constrain some evolutionary scenarios between these types of objects.
The type-2 quasars vary considerably in CO mass. The two $z\sim 3.5$ 
objects from Polletta et al. (2011) have gas masses comparable to 
SMGs, but the $z\sim 2.8$ objects from this paper and the $z=2.8$
and $z=4.1$ objects from 
Martinez-Sansigre et al.\ (2009) have detections or limits significantly 
lower than this. With the exception of XFLS171419.97+602724.8, the type-2
quasars tend to have similar stellar masses to the SMGs.

Making the assumption
of conversion of gas to stars with 100\% efficiency,
trajectories can be drawn on Figure 6 (such as the large
magenta arrow) to indicate how a galaxy might evolve with time.
(Although such a scenario is highly simplistic as feedback effects will
prevent the complete conversion of gas to stars, more gas is likely
to accumulate too, both from expelled gas cooling and infalling, and
from accretion of fresh material.) 
This simple scenario would indicate that 
SMGs and CO-rich type-2 quasars could evolve into 
moderately-high stellar mass galaxies, comparable in mass to 
the hosts of normal type-2 quasars. However, unless they are able
to accrete significantly more gas than is currently present in their
molecular reservoirs (or merge with other massive galaxies), 
they would not be able to evolve
into our most massive host galaxy , XFLS171419.97+602724.8. 
It may be significant that, although 
this object is extremely luminous in the infrared, it 
has a high dust temperature, perhaps 
indicating a more evolved system overall, with AGN heating 
dominating the infrared SED.
The $z\sim 2$ ULIRGs seem to be a distinct population of relatively low
stellar mass objects that could evolve into some of the lower 
mass (and lower luminosity) type-2 quasars. Similarly, the BX and 
BzK galaxies also look to be distinct from the majority of the SMGs
and type-2 quasars. Clearly, however, these is a need for larger samples
of high redshift galaxies with better controlled selection effects. 
These can now be provided using the new generation of micro-Jansky depth, 
$\sim 10-20$deg$^2$ near-infrared surveys such as the {\em Spitzer} 
Extragalactic Representative Volume Survey (Mauduit et 
al.\ 2011) in combination with far-infrared surveys with Herschel
(e.g.\ Oliver et al.\ 2010) and 
follow-up with EVLA and the Atacama Large mm/submm Array (ALMA).

\acknowledgments

We thank Chris Carilli for helpful advice on EVLA observations, 
Hana Schumacher for providing a more accurate estimate of the CO (1-0)
luminosity for one of the high redshift radio-intermediate type-2 quasars, 
and the referee for comments that significantly improved the manuscript. 
The National Radio Astronomy Observatory is a facility of the National Science Foundation, 
operated under cooperative agreement by Associated Universities, Inc.
Observations reported here were obtained at the MMT Observatory, a joint facility 
of the Smithsonian Institution and the University of Arizona. MMT time was granted by 
NOAO through the Telescope System Instrumentation Program (TSIP). TSIP is funded by NSF.
ML and AP were visiting astronomers at the Infrared Telescope Facility, which is 
operated by the University of Hawaii under Cooperative Agreement no. NNX-08AE38A with 
National Aeronautics and Space Administration, Science Mission Directorate, Planetary
Astronomy Program. The NASA/IPAC Extragalactic Database (NED) and Infrared
Science Archive (IRSA) are operated by the Jet 
Propulsion Laboratory, California Institute of Technology, under contract with the National
Aeronautics and Space Administration. {\em Herschel} is an ESA space observatory with science
instruments provided by European-led Principal Investigator consortia with significant
participation from NASA.

\end{document}